\documentclass[prb,twocolumn,a4paper,showpacs,floatfix]{revtex4}
\usepackage{graphicx}
\usepackage{amsmath}

\begin{document}

\title{The connection between noise and quantum correlations in a
double quantum dot}

\author{F. Bodoky}
\affiliation{Kavli Institute of NanoScience, Delft University of
 Technology, Lorentzweg 1, 2628 CJ Delft, The Netherlands}
\author{W. Belzig}
\affiliation{Fachbereich Physik, Universit\"at Konstanz, 78457
  Konstanz, Germany}
\author{C. Bruder}
\affiliation{Departement of Physics, University of Basel, Klingelbergstr. 82, 
4056 Basel, Switzerland}
\date{\today}
\begin{abstract}
  We investigate the current and noise characteristics of a double
  quantum dot system. The strong correlations induced by the Coulomb
  interaction and the Pauli principle create entangled two-electron
  states and lead to signatures in the transport properties. We show
  that the interaction parameter $\phi$, which measures the admixture
  of the double-occupancy contribution to the singlet state and thus
  the degree of entanglement, can be directly accessed through the
  Fano factor of super-Poissonian shot noise.
\end{abstract}



\maketitle

\section{Introduction}

Quantum dots, their current and noise properties
\cite{blanter:00,nazarov:03} are widely investigated topics in today's
nanophysics with a number of possible applications, the most
revolutionary being a spin-based quantum computer
\cite{LossDiVicenzo}. Here, the spin of an electron in a quantum dot
is used as a qubit, which serves as the basic building block of a
quantum computer. A necessary ingredient for quantum processing is the
possibility to couple at least two qubits and to understand and
manipulate the various correlations.  This is one of the motivations
to study double quantum dots
\cite{bur99,ziegler00,wie03,golovach2003,gol04,gra06,bulka:04,dicarlo:04,hatano:04,wunsch:06}.
In particular, it is required to create entangled electron states by
the interaction of an electron inside the double dot with an electron
tunneling onto the double dot. Measuring this entanglement is an
important experimental task and theoretical suggestions how to probe
these states are needed.

In this work, we will discuss the relation between zero-frequency
current noise, Coulomb correlations, and entanglement for the example
of a double quantum dot. We will make use of the recent observation
\cite{cottet,belzig2005,cottet:prb} that noise measurements in the nonlinear
Coulomb-blockade regime can be used to obtain spectroscopic
information of excited levels in a quantum dot. This scheme will be
adapted to a double quantum dot and we will show that the important
interaction parameter $\phi$ (containing information about the 
entanglement)~\cite{golovach2003,gol04} enters the Fano
factor (i.e., the ratio of noise power to current) in this
regime. This kind of noise spectroscopy can be used to directly
extract $\phi$ from experimentally observable quantities. We extend a
previous discussion \cite{aghassi:06} of the low-frequency shot noise
of a double-dot system by an analysis of the full counting statistics
of the transferred charge. This allows us to identify directly the
interaction parameter by a noise measurement. One main result is
obtained in the transport regime, in which predominantly only one 
electron occupies the lowest double dot level due to the Coulomb
interaction. Thermal activation leads to tunneling events through
excited two-electron states which result (under the conditions
discussed below) in super-Poissonian shot noise characterized by a
Fano factor
\begin{equation}\label{eFanC}
 F(\phi) = 1 + \phi^2\,.
\end{equation}
Hence, a measurement of the noise in this regime
allows a direct determination of the interaction parameter $\phi$. 

\section{Model and Theoretical Methods}

We now turn to a description of the model and the method. 
With maximally two electrons occupying the double dot and no magnetic 
field lifting the spin-degeneracy, the following
Hund-Mulliken eigenstates of the double quantum dot are possible (for
a more detailed description see \cite{bur99,gol04}): the zero-electron
state $|0\rangle$, two two-fold degenerate one-electron states
$|+\rangle$ (symmetric) and $|-\rangle$ (anti-symmetric), four
two-electron states with one electron per dot, the singlet $|S\rangle$
and the three-fold degenerate triplet $|T\rangle$. There are two other
singlet states that have a significantly higher energy due to the
large on-site Coulomb interaction which do not have a physical 
effect in the regime studied (but are included in the numerical 
calculations).

We consider a (longitudinal) double dot coupled to leads, which are
modeled as a Fermi sea.  The tunnel coupling leads to a
tunneling amplitude $t_\alpha$, which measures the overlap between the
orbital state and the lead wave function in terminal $\alpha$.  We
assume that $V$ is the applied bias voltage and that $\mu_L= eV/2$
($\mu_R= -eV/2$) are the
chemical potentials of the left and right leads. There is an
additional lead with capacitive coupling to the double dot
characterized by the gate voltage $V_{g}$ that shifts the potential on
the dot. When one electron is on the dot, it can be in two possible
states: its wave function is spread either symmetrically or
anti-symmetrically over the two dots. The symmetric state has a lower
eigenenergy, and the energy difference between the two one-electron
states depends on the interdot tunneling amplitude $t_{0}$, describing
the potential barrier between the two dots. Including the energy
contribution of the gate voltage, the eigenenergies of these states
can be calculated as $E_{\pm} = eV_{g} \mp t_{0}$. Adding a second
electron, we have first to overcome the gate voltage again, and
additionally the repulsive Coulomb interaction between the two
electrons leads to a spin-dependent splitting of the eigenenergies.
For the singlet state $|S\rangle$ we obtain the energy $E_{S} =
2eV_{g} + u_{12} - J$, and for the (three-fold degenerate) triplet
state $|T \rangle$ the energy $E_{T} = 2eV_{g} + u_{12}$. Here,
$u_{12}$ is the (inter-dot) electron-electron repulsion and
$J=4t_0^2/u_H$ is the Heisenberg exchange parameter that characterizes
the Heisenberg interaction between the two spins, $H_{spin} = J
\mathbf{S}_{1}\cdot\mathbf{S}_{2}$; $u_H$ is the on-site Coulomb
repulsion. Because $J$ is positive, the singlet state has a lower
energy than the triplet state. 

The singlet state can be expressed in terms of the creation operators
$d_{n}^{\dagger}$, $n = \pm$, which create an (anti)-symmetric
electron state 
\begin{equation}\label{eSinglet}
 | S \rangle = \frac{1}
{\sqrt{1+\phi^{2}}}\left(d_{+\uparrow}^{\dagger}d_{+\downarrow}^{\dagger} - 
\phi d_{-\uparrow}^{\dagger}d_{-\downarrow}^{\dagger} \right) | 0 \rangle\,.
\end{equation}
The interaction parameter $\phi$ describes the competition between the
kinetic-energy gain and the Coulomb repulsion. In the Hund-Mulliken
model $\phi = \sqrt{1 + 16t_{0}^{2}/u_H^2} - 4t_{0}/u_H$ is
determined by the inter-dot tunneling amplitude $t_{0}$ and the
on-site Coulomb repulsion $u_H$ \cite{bur99}.  In general, the
dependence of $\phi$ on the microscopic parameters may have a
different form, however we will always use this definition for the
quantitative plots below.  It is important to stress that the results
obtained below are largely independent on the precise dependence of
$\phi$ on the details of the quantum dot. Hence, the noise features we
find can be used as an additional test of the applicability of the
model calculation of Ref.~\cite{bur99} to realistic quantum dots.

We describe transport through the double dot in the 
sequential-tunneling approximation using the Master equation
\begin{equation}\label{eMastEq}
 \frac{d P_{i}}{d t} = 
 \sum_{j}\left(\Gamma_{ji}P_{j} - \Gamma_{ij}P_{i}\right)\,.
\end{equation}
Here, $i$ and $j$ label the available states of the quantum dot,
including the number of electrons on the dot, the orbital index, and
the spin. $P_{i}$ is the occupation probability of state $i$, and
$\Gamma_{ij}$ is the tunneling rate from state $i$ to state $j$. 
The rates are given by 
\begin{eqnarray}\label{eGoldRule}
 \Gamma_{ji} & = & \gamma_Lm_{ij}f(\sigma_{ij}(\epsilon_{ij}-\mu_{L})) + 
 \gamma_Rm_{ij}f(\sigma_{ij}(\epsilon_{ij}-\mu_{R})),
\nonumber
\end{eqnarray}
Here, $\gamma_{L/R}=\nu_{L/R}|t_{L/R}|^2/\pi$ is the bare tunneling
rate to the left (right) terminal, and $\nu_{L/R}$ is the density of
states in the leads. The matrix elements $m_{ij}$ depend on the
orbital and the spin state of the dot \cite{gol04}. The occupation in
the leads is determined by the Fermi functions $f(\epsilon) =
1/[\exp(\epsilon/k_\mathrm{B}T)+1]$, the direction of tunneling is
defined by $\sigma_{ij}=\pm 1$ and the energy difference by
$\epsilon_{ij} = E_{i} - E_{j}$.  We note here, that the interaction
parameter $\phi$ enters the calculations through the matrix elements
$m_{ij}$ of the singlet state. For a more detailed discussion of these
tunneling rates we refer to \cite{gol04}.  For later use we express
the Master equation~(\ref{eMastEq}) in matrix form, $d\mathbf{P}/dt =
\mathbf{MP}$, where $M_{ij} = \Gamma_{ji}$ for $i\neq j$ and $M_{ii} =
-\sum_{j\neq i}\Gamma_{ij}$.

From the Master equation, we obtain the noise
\cite{korotkov,hershfield,hanke,bagrets2003,bulka}. The quantity 
characterizing the current noise is the Fano factor $F = S/2eI$. It
measures, how far the distribution of the tunneling events differ from
the random, Poissonian distribution: for Poissonian noise, when the
tunneling events are uncorrelated, $F=1$. A Fano factor $F<1$ can be
due to anti-correlations (e.g. the Pauli principle). We will be mostly
interested below in super-Poissonian noise, $F>1$, and we will show
that here it is due to correlated transport cycles resulting from a
blockade of open channels.  By a transport cycle we mean the following
process \cite{belzig2005}: suppose the system is initially in its
ground state $A$, and another state $B$ (which would be available for
tunneling according to its energy) is blocked due to Coulomb
interaction. When this electron in $A$ tunnels out of the dot due to a
thermal excitation, both the states $A$ and $B$ are available for
electrons to tunnel into. A cycle is occurring, when a sequence of
electrons tunnels through state $B$ before an electron enters the
ground state $A$ and blocks the transport through the dot again. This
cycle leads to a correlated transfer of a number of charges given by
the number electrons tunneling through the excited state $B$. Hence,
that effective charge transferred in one cycle is larger than one and
leads to an increased Fano factor.

The above picture is based on the sequential-tunneling
approximation. The parameters have to be chosen such that cotunneling
processes can be neglected. This is experimentally possible, since the
current due to cotunneling processes is generally of the order of
$t^4/\Delta $,
where $t$ is the tunneling amplitude and $\Delta$ is the excitation
energy in the virtual intermediate state. In our case the current in
the most interesting regime is proportional to $t^2x$, where $x$
is exponentially small in $\Delta/k_{\mathrm{B}}T$.  Since the
tunneling amplitude $t$ can in principle be made arbitrarily small,
sequential tunneling becomes the dominant process. Recently it was
shown that noise spectroscopy deep inside the Coulomb blockade regime
is possible \cite{ensslin2006,fujisawa2006,ensslin2007,dicarlo:07} and in
agreement with the sequential-tunneling description.

\section{Current and Fano factor}

\begin{figure}[ht]
 \begin{center}
 \scalebox{0.85}{\includegraphics{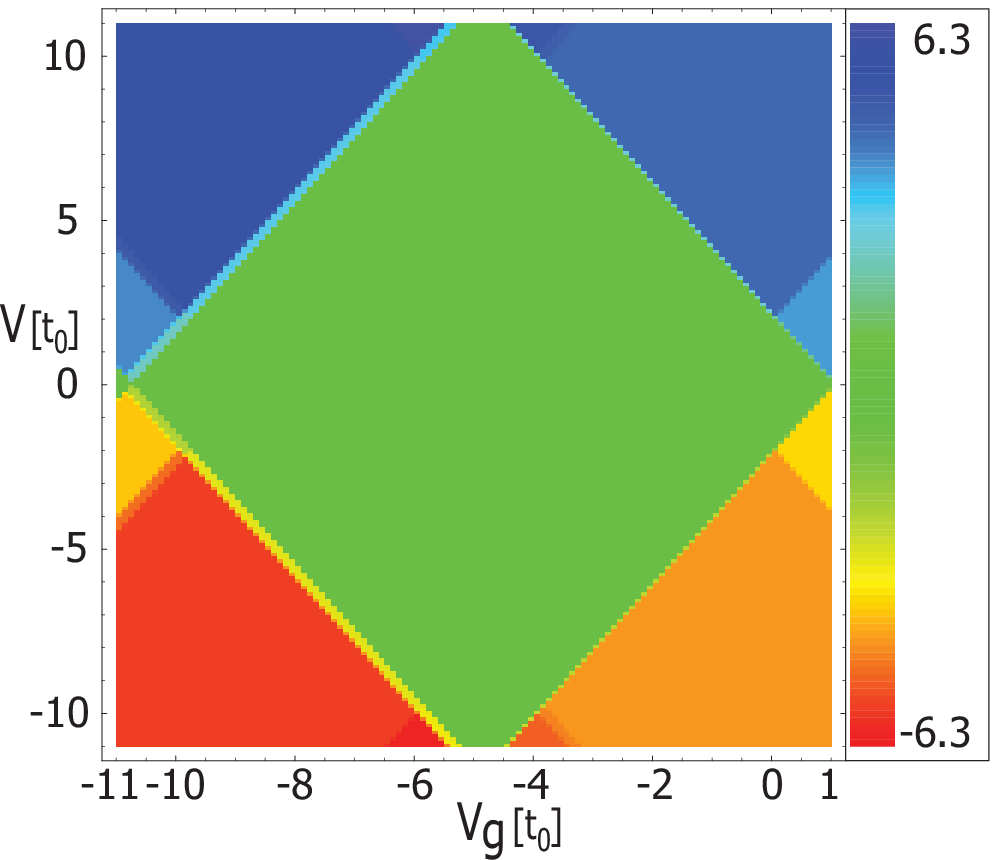}}
 \scalebox{0.85}{\includegraphics{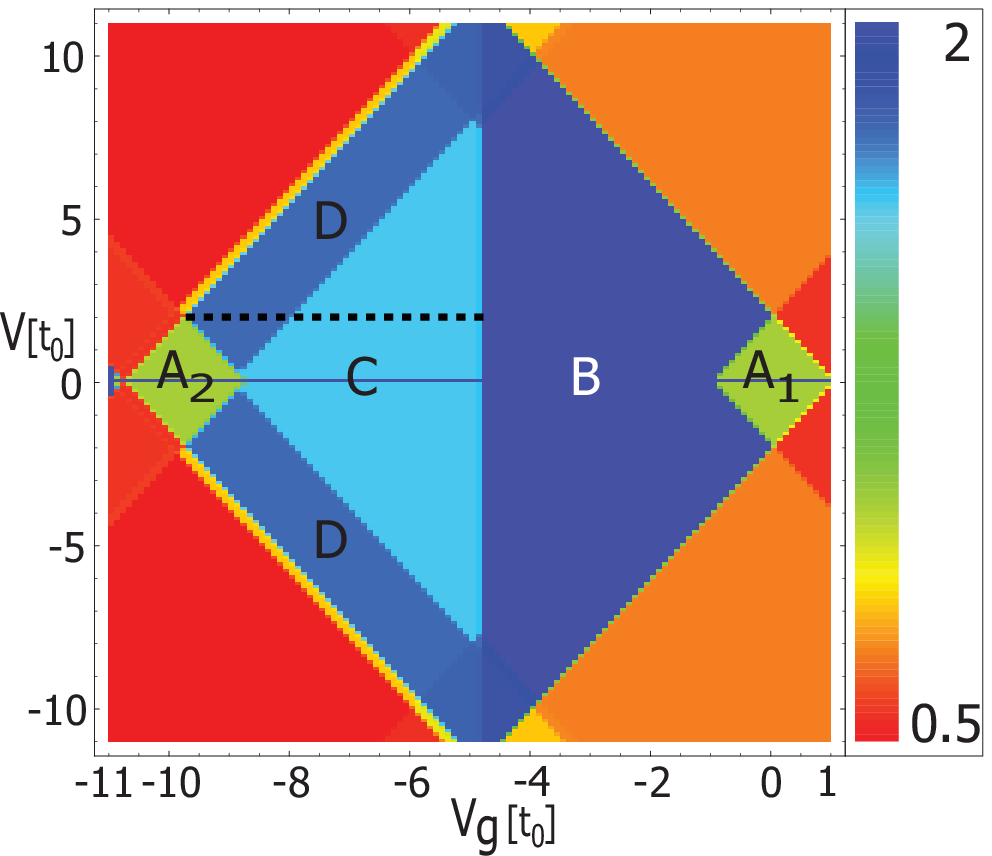}}
 \caption{(Color online) Current (top) and Fano factor (bottom) in
  the one-electron blockade regime as a function of gate voltage
  $V_{g}$ and bias voltage $V$. The current is given in units of
  $e\gamma$. 
  Note the various plateaus in the Fano factor within the blockade
  regime, which are invisible in the current. The letters label the
  subregions as referred to in the main text. The dashed black line
  in the bottom plot indicates the line along which Figs.~\ref{fPhiDep}
  and~\ref{fig:temp} are plotted. In this plot we use the parameters
  $u_H=17t_{0}$, $u_{12}=10t_{0}$, $\gamma_L=\gamma_R=\gamma$ and
  $k_BT=0.01t_0$ (and hence $J=0.24t_0$ and $\phi=0.79$).}
 \label{fBlocCurFan}
 \end{center}
\end{figure}

The results of our calculation are summarized in
Fig.~\ref{fBlocCurFan}, which shows the average current (top panel)
and the Fano factor (bottom panel) as a function of the bias voltage
$V$ and the gate voltage $V_{g}$.  Considering the current, the most
visible feature is the Coulomb blockade diamond (green in
Fig.~\ref{fBlocCurFan}). As long as the bias voltage is lower than the
energy difference between the energetically lowest state and the next
state with one electron more or less, i.e., $eV < E_{i}(V_{g}) -
E_{j}(V_{g})$ with $i$ and $j$ being two states differing by one
electron, electron tunneling is exponentially suppressed. In our
model, we have three such blockade regimes with 0, 1, and 2 electrons
in the dot for zero bias voltage. The two parts of Fig.~\ref{fBlocCurFan} show
the current and the Fano factor for the most interesting of these
blockade regions, the one-electron blockade regime. We will
concentrate our discussion on this region, since most interesting
features can be discussed here. 

\begin{figure}[ht]
  \begin{center}
    \scalebox{0.85}{\includegraphics{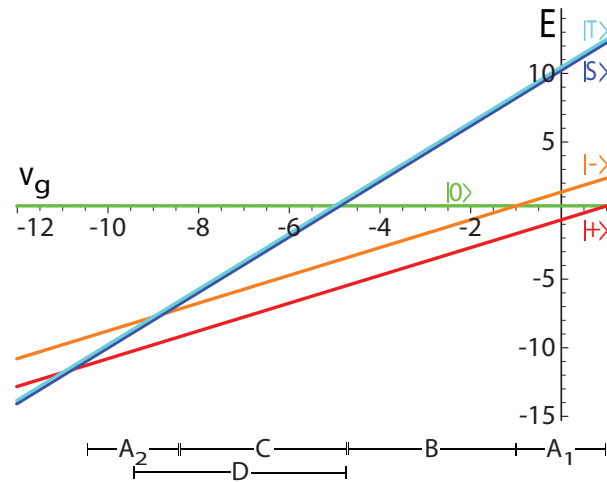}}
    \caption{(Color online) Energy of the states as a function of gate voltage
      $V_{g}$ in units of $t_{0}$. The capital letters correspond 
      to the regions in Fig.~\ref{fBlocCurFan}. 
      The parameters are chosen like in
      Fig.~\ref{fBlocCurFan}. 
    }\label{fEnPlot}
  \end{center}
\end{figure}

\begin{figure}[ht]
 \begin{center}
   \scalebox{0.85}{\includegraphics{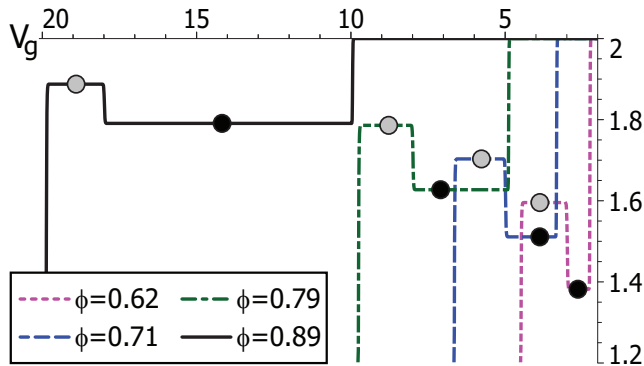}}
   \caption{(Color online) Fano factor for different values of $\phi$ in the regions
     $C$ and $D$ as a function of gate voltage (in units of $t_0/e$). The bias 
     voltage is $eV = 2t_{0}$ and $k_BT=0.01 t_0$. Furthermore, the ratio $u_{12}/u_{H} 
     = 0.6$ is kept fixed, such that the different plots can be easily achieved in experiment 
     by changing the interdot tunneling rate $t_{0}$. The parameters are chosen as 
     $(\phi;\, J/t_{0},\, u_{12}/t_{0})=(0.62;0.5,5),\, 
     (0.71;0.34,7),\, (0.79;0.24,10),$ $(0.89;0.12,20)$ from right to left. The 
     plateau heights, indicated by the black (region C) and gray (region D) dots, 
     follow from the analytic expressions Eqs.~\eqref{eFanC} and~\eqref{eFanD}. }
 \label{fPhiDep}
 \end{center}
\end{figure}

We first look at the energies of the states in the blockade regime as
a function of the gate voltage, see Fig.~\ref{fEnPlot}. The capital
letters in Figs.~\ref{fBlocCurFan} and~\ref{fEnPlot} correspond to
these blockade regions, and we will now show that the features in the
Fano factor can be explained by the energy plot for the states in this
region. We first consider the regions $A$. They are both characterized
by a Fano factor $F=1$, typical tunneling processes through a simple
two-level system.  Indeed, Fig.~\ref{fEnPlot} shows that the first
excited states differ in electron number.  In $A_{1}$ the excited
state is $|0\rangle$ and in $A_{2}$ $|S\rangle$. Lowering the gate
voltage, we come to region $B$, where the first excited state is the
state $|-\rangle$, i.e., a state with one electron as well. Thus, the
description in terms of a single state, which can be occupied or not,
does not hold any more, and we observe an increase in the Fano factor
to $F=2$. In this region, the state $|0\rangle$ is still the first
state available for tunneling, but the blocked state $|-\rangle$ is
energetically lower. Continuing to region $C$, in which the energies
of the two-electron states become lower than the zero-electron
state. So the first available excited state is the singlet
$|S\rangle$.  The difference between the regions $C$ and $D$ is
somewhat more complicated. Since the difference $E_{S} - E_{-}$ is
smaller than $E_{S} - E_{+}$, the state $|+\rangle$ still blocks the
dot, if it is occupied. After a thermal excitation to the state
$|S\rangle$, electrons tunnel through the states $|S\rangle$ and
$|-\rangle$ until an electron occupies the state $|+\rangle$ again and
blocks the dot. In the regions $C$ and $D$ the Fano factor depends on
the parameter $\phi$ through the prefactor of the singlet states (see
Eq.~\eqref{eSinglet}). The $\phi$-dependence is illustrated in
Fig.~\ref{fPhiDep} by varying the parameter $u_H/t_0$. As can be seen
the positions of the steps change simultaneously with the plateau
height. However, the height depends only on the interaction parameter,
as we have checked by numerically varying $u_H$ and $\phi$
independently. We will derive below analytic expressions for the Fano
factor, which confirm this behavior.

\begin{figure}[ht]
 \begin{center}
 \scalebox{0.85}{\includegraphics{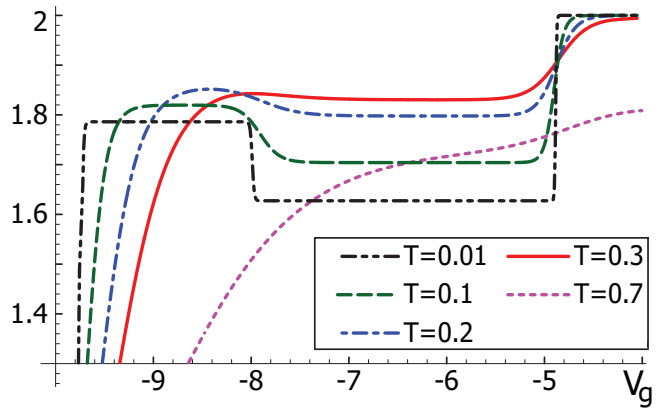}}
 \caption{(Color online) Fano factor as a function of gate voltage (in units of
   $t_{0}/e$) at various temperatures for $\phi = 0.79$, $u_H=17t_{0}$, and 
   $u_{12}=10t_{0}$. The steps in the Fano factor are washed out by the temperature, 
   however the noise remains super-Poissonian in the whole regime shown.
 }
 \label{fig:temp}
 \end{center}
\end{figure}

The effect of temperature will be to wash out the sharp steps and
plateaus. This is illustrated in Fig.~\ref{fig:temp}. Increasing the
temperature to a value of the order of the level splitting first the
sharp steps are washed out. The plateaus C and D, however, remains
still visible and the values are still given, at least approximately,
by the low temperature results. Finally, the temperature leads to a
vanishing of the plateaus, which make it difficult to extract the
interaction parameter $\phi$ from the Fano factor.

\section{Counting Statistics}

A more detailed view of the transport characteristic can be obtained
by looking at the full counting statistics (FCS)
\cite{bagrets2003}.  The aim is here to find the cumulant
generating function (CGF) $\mathcal{S}(\chi)$, which is related to the
probability $P(N)$ of $N$ charges passing through the system during
the measurement time $t_M$ by $\exp\mathcal{S}(\chi) = \sum_{N}P(N)
e^{iN\chi}$. The knowledge of $\mathcal{S}(\chi)$ is equivalent to the
knowledge of all the cumulants of the system according to
$C_k=\left.(-i\partial/\partial_{\chi})^k\mathcal{S}(\chi)\right|_{\chi=0}$,
such as current (first cumulant) and noise (second cumulant),
$I=eC_{1}/t_M$ and $S = 2e^{2}C_{2}/t_M$, respectively (where $e$ is
the electron charge). The simplest way to evaluate the counting
statistics in this case is to use the method described in
Ref.~\onlinecite{bagrets2003}. We choose (without loss of generality)
to count the charges in the left lead. Therefore, we have to change
the left tunneling rates in the off-diagonal elements of the transport
matrix $\mathbf{M}$: $\Gamma_{L} \rightarrow
\Gamma_{L}\exp(i\chi)$. The smallest eigenvalue of $\mathbf{M}$, which
we denote by $\lambda_{0}(\chi)$, determines the CGF as
$\mathcal{S}(\chi) = -t_M\lambda_{0}(\chi)$.

If we restrict our calculations to a certain region and thus reduce
the number of involved states, it is possible to get an analytical
expression for the Fano factor \cite{belzig2005}. For region $C$, we
include the one-electron states $|+\rangle$, $|-\rangle$, and the
two-electron singlet $|S\rangle$. Here, the relevant energy
differences are both smaller than the bias: $E_{S}-E_{-} < E_{S}-E_{+}
< eV/2$.  
Therefore both Fermi functions for the tunneling processes
are exponentially suppressed, but one is much bigger than the other:
\begin{equation}\label{eApproxC}
 1 \gg f(E_{S}-E_{+}-eV/2) \gg f(E_{S} - E_{-} - eV/2).
\end{equation}
Therefore, we neglect $ f(E_{S} - E_{-} - eV/2)$ and take the rates to lowest
order in the parameter $x\equiv f(E_{S}-E_{+}-eV/2)$. Then the tunneling
rates are
\begin{eqnarray}\label{eTransC}
 \Gamma_{+S} \approx \frac{1}{1+\phi^{2}}\gamma_{L}x, & \qquad &
 \Gamma_{S+} \approx \frac{2}{1+\phi^{2}}(\gamma_{L}+\gamma_{R}), \nonumber\\
 \Gamma_{-S} \approx \frac{\phi^{2}}{1+\phi^{2}}\gamma_{L} &, \qquad &
 \Gamma_{S-} \approx \frac{2\phi^{2}}{1+\phi^{2}}(\gamma_{L}+\gamma_{R}).
 \nonumber
\end{eqnarray}
Here, $\gamma_{L/R}$ are the bare tunneling rates for the left/right
leads. Without loss of generality, we will count the charges in the
left lead, which means, we have to replace
$\gamma_{L}\leftrightarrow\gamma_{L}\exp(\pm i \chi)$ (the sign
depending on whether an electron enters or leaves the dot
respectively) in the off-diagonal elements of the matrix $\mathbf{M}$.
The counting statistics is obtained from the smallest eigenvalue
$\lambda_{0}$, which we determine to lowest order in $x$.
The result is
\begin{equation}
 \label{eq:fcs_C}
 S(\chi)=-x\frac{\gamma_L\gamma_R(1+\phi^2)}{\gamma_R(1+\phi^2)-\gamma_L}
 \frac{e^{i\chi}-1}{1-p(\phi) e^{i\chi}}\,,
\end{equation}
where $p(\phi)=\phi^2\gamma_R/[\gamma_L+\gamma_R(1+\phi^2)]$.
Finally we obtain the Fano factor
\begin{equation}
 F(\phi)=\frac{1+p(\phi)}{1-p(\phi)}=
1+\phi^2\frac{2\gamma_R}{\gamma_L+\gamma_R}\,,
\end{equation}
which is independent of the bias voltage as long as we are in the 
one-electron Coulomb blockade regime. For a symmetric structure with
$\gamma_L=\gamma_R$ we obtain Eq.~(\ref{eFanC}).

A similar result may be obtained for the region $D$ in
Fig.~\ref{fBlocCurFan}. Here, the situation is a bit more complicated,
since the lowest-lying triplet state also plays a role and, thus, the
problem involves four states. In the same way as before we neglect
$f(E_{T} - E_{\pm} - eV/2)$ since the transition
between $|S\rangle\leftrightarrow|+\rangle$ is the least
suppressed. Assuming that the temperature is much smaller
than the singlet-triplet splitting, we calculate the series expansion
for the Fano factor (with symmetric leads) to be
\begin{equation}\label{eFanD}
 F(\phi) = \frac{2 + 17\phi^{2} + 32\phi^{4} + 16\phi^{6}}{2 + 
11\phi^{2} + 16\phi^{4} + 3\phi^{6}}\,.
\end{equation}
Both these results, Eqs.~(\ref{eFanC}) and (\ref{eFanD}), are in
agreement with our numerical calculations as shown in Fig.~\ref{fPhiDep}.

\section{Conclusions}

In conclusion, we have shown that noise measurements in the Coulomb
blockade regime of a double quantum dot can reveal interesting
information about the quantum correlations created during
tunneling. In particular, we have shown that the Fano factor in the
one-electron blockade region is super-Poissonian and can be used to
determine the interaction parameter $\phi$. This allows to measure the
degree of entanglement of two electrons in the double quantum dot.

We would like to thank D. Loss for useful discussions. This
work was financially supported by the Swiss National Science
Foundation, the NCCR Nanoscience,
by the Deutsche Forschungsgemeinschaft within the SFB 513 and the
Priority Program Semiconductor Spintronics and by the Landesstiftung
Baden-W\"urttemberg within the Kompetenznetzwerk Funktionelle
Nanostrukturen.

\end{document}